\documentclass[twocolumn,showpacs,preprintnumbers,amsmath,amssymb,prb]{revtex4}

\usepackage{graphicx}
\usepackage{color}

\begin{document}

\title{Thermodynamic properties of graphene bilayers}
\author{Carlos P. Herrero}
\author{Rafael Ram\'irez}
\affiliation{Instituto de Ciencia de Materiales de Madrid,
         Consejo Superior de Investigaciones Cient\'ificas (CSIC),
         Campus de Cantoblanco, 28049 Madrid, Spain }
\date{\today}

\begin{abstract}
Thermodynamic properties of graphene bilayers are studied by 
path-integral molecular dynamics (PIMD) simulations, 
considering quantization of vibrational modes and
anharmonic effects. 
Bilayer graphene has been studied at temperatures between 
12 and 1500~K for zero external stress, using the LCBOPII 
effective potential.
We concentrate on the thermal expansion, in-plane and 
out-of-plane compressibility, and specific heat.
Additional insight into the meaning of our results for bilayer
graphene is obtained from a comparison with data obtained from
PIMD simulations for monolayer graphene and graphite. They are also
analyzed in view of experimental data for graphite.
Zero-point and thermal effects on the in-plane and ``real'' area 
of bilayer graphene are studied. The thermal expansion coefficient 
$\alpha_{xy}$ of the in-plane area is negative at low temperatures and 
positive for $T \gtrsim$ 800~K. 
The minimum $\alpha_{xy}$ is $-6.6 \times 10^{-6}$ K$^{-1}$
at $T \approx 220$~K.
Both in-plane ($\chi_{xy}$) and out-of-plane ($\chi_z$) compressibilities 
of graphene bilayers
are found to increase for rising temperature, and turn out to be lower 
than that corresponding to monolayer graphene and higher than 
those found for graphite.
At 300 K, we find for the bilayer $\chi_{xy} = 9.5 \times 10^{-2}$ \AA$^2$/eV 
and $\chi_z = 2.97 \times 10^{-2}$ GPa$^{-1}$.
Results for the specific heat obtained from the simulations are 
compared with those given by a harmonic approximation
for the vibrational modes. This approach is noticeably accurate 
at temperatures lower than 200~K.
\end{abstract}


% 61.48.Gh, Graphene structure
% 65.80.Ck, Thermal properties of graphene
% 63.22.Rc, Phonons in graphene

\maketitle

\section{Introduction}

Graphene bilayers have attracted great interest in last years
after the finding that they present unconventional superconductivity
when stacking both sheets twisted relative to each other
by a small angle.\cite{ca18,gu18,ya19}
It has been also recently noticed the existence of Mott-like insulator
states in these materials, for the appearance of localized electrons
in the superlattice corresponding to a moir\'e pattern.\cite{ca18b,po18}
Moreover, twisted graphene bilayers display magnetic properties
which can be externally controlled by an applied
bias voltage.\cite{go17,sb18}
Graphene bilayers are known to show ripples and out-of-plane
deformations similar to the monolayers,\cite{me07b}
causing a departure from planarity which is believed to
be a relevant mechanism for electron scattering.\cite{gi10}

From a basic point of view, understanding the thermodynamic 
properties of two-dimensional (2D) systems in
three-dimensional (3D) space has been along the years a continuous
objective in the field of statistical physics.\cite{sa94,ne04}
This problem has been mainly treated in connection with
soft condensed matter and biological membranes,\cite{ch15,ru12}
whose complexity makes it very demanding to devise
microscopic models built on realistic interatomic interactions.
Graphene bilayers are a well-controlled instance of
crystalline membranes formed by two atomic sheets, for which
an atomic-level description is possible, allowing for
a deep insight into the physical properties of this kind
of systems.\cite{ba11,al13,al14,am14,ma16}
Moreover, graphene shows us as a suitable material to study
the thermodynamic stability of 2D crystals, which has been long
discussed and can be related to anharmonic coupling between
in-plane and out-of-plane vibrational modes.\cite{am14,co16}

Various kinds of atomistic simulations have been employed
to study finite-temperature properties of
graphene.\cite{fa07,ca09c,ak12,ma14,lo16}
In most of them, C atoms were considered as classical particles,
but the Debye temperature of graphene for out-of-plane vibrational
modes is $\Theta_D^{\rm out} \gtrsim$ 1000~K and higher
for in-plane modes.\cite{po11}
This indicates that the influence of quantum fluctuations on
physical properties should be appreciable even for $T$ much higher
than room temperature.

Some works have presented path-integral-type simulations,
which allow one to study thermal and quantum fluctuations at
finite temperatures.
This type of simulations have been carried out for graphene monolayers
to study structural and thermodynamic properties
of this material.\cite{br15,he16,he18,ha18}
In addition to this, nuclear quantum effects have been analyzed
earlier by means of a combination of density-functional theory and
a quasi-harmonic approximation for vibrational modes
in this crystalline membrane.\cite{mo05,sh12}

The thermal behavior of monolayer graphene has been studied
by means of path-integral simulations,\cite{he18}
with particular emphasis on low temperatures.
In this paper we extend that analysis to graphene bilayers,
where new aspects are expected to appear due to interlayer
interactions, and the associated coupling between atomic
displacements of both layers in the out-of-plane direction.

We employ the path-integral molecular dynamics (PIMD)
method to study thermodynamic properties of graphene bilayers at
temperatures between 12 and 1500~K.
Simulation cells of different sizes are considered, as finite-size effects
have been found earlier to be important for some equilibrium
properties of graphene.\cite{ga14,lo16,he16}
We analyze the thermal behavior of the sheet surface
in graphene bilayers, considering
the difference between {\em real} and {\em in-plane} area.
We put special attention on the temperature dependence of
the thermal expansion, compressibility, and specific heat $c_p$.
In particular, low-temperature results of the simulations for $c_p$
are compared with the prediction of a harmonic approximation for
the vibrational modes.

 The paper is organized as follows. In Sec.\,II we describe the
computational method used in the simulations.
In Sec.~III we present results for the real and in-plane areas,
as well for the so-called {\em excess} area of graphene bilayers.
The thermal expansion is discussed in Sec.~IV, and the
compressibility (in-plane and out-of-plane) is analyzed in Sec.~V.
In Sec.~VI we present results for the specific
heat, and in Sec.\,VII we summarize the main results.

\section{Computational Method}

\subsection{Path-integral molecular dynamics}

Here we employ PIMD simulations to study structural and thermodynamic
properties of graphene bilayers as a function of temperature.
This method, based on the Feynman path-integral formulation of statistical
mechanics,\cite{fe72} is now a well-established nonperturbative approach
to investigate finite-temperature properties of many-body quantum systems.
In the applications of this computational technique to numerical 
simulations, each quantum particle (here atomic nucleus) is represented
as a group of $N_{\rm Tr}$ beads (the so-called {\em Trotter number}), 
behaving like classical particles disposed to form a ring 
polymer.\cite{gi88,ce95,he14}

In actual simulations of condensed matter using the path-integral
method, the configuration space of the classical isomorph is explored
by means of molecular dynamics or Monte Carlo sampling.
In this paper we use molecular dynamics, as we have found that our
computing codes are more effectively parallelizable with this
procedure.  We note that the dynamics in this kind of 
PIMD simulations is artificial, in the sense that it does not 
reproduce the dynamics of the actual quantum particles
under consideration. Nevertheless, it is very efficient to
sample the many-body configuration space, giving precise results for
time-independent equilibrium properties of the quantum system.

We describe the interatomic interactions in graphene with a 
long-range carbon bond-order potential, 
the so-called LCBOPII,\cite{lo05}
which has been employed earlier to carry out classical simulations 
of carbon-based systems, such as diamond,\cite{lo05} 
graphite,\cite{lo05}, and liquid carbon.\cite{gh05}
It has been more recently applied to study 
graphene,\cite{fa07,za10b,lo16} with particular emphasis on its 
mechanical properties.\cite{za09,ra17}
The LCBOPII potential has been also used to perform PIMD simulations
of graphene monolayers\cite{he16} and bilayers,\cite{he19}
which has allowed an assessment of quantum effects
by comparing with results of classical simulations.
In this paper, according to earlier simulations,\cite{ra16,he16,ra17}
the original LCBOPII parameterization has been slightly changed 
to increase the zero-temperature bending constant 
$\kappa$ of a graphene monolayer from 0.82 eV to a more realistic 
value of 1.49 eV, closer to experimental data and
{\em ab-initio} calculations.\cite{la14}
The interlayer interaction is the same as that employed in earlier
simulations of bilayer graphene with this effective 
potential.\cite{za10b,he19}
Thus, the interlayer binding energy for the minimum-energy 
configuration with AB stacking is 25 meV/atom for bilayer
graphene and 50 meV/atom for graphite.

Our simulations of graphene bilayers have been performed in the 
isothermal-isobaric ensemble, where we fix the number of carbon 
atoms ($2N$), the in-plane stress (here $P_{xy} = 0$), 
and the temperature ($T$).
We employed effective algorithms for carrying out PIMD simulations, 
as those presented in the literature.\cite{tu98,ma99} 
Specifically, staging variables\cite{tu93} were used to define the 
bead coordinates, and a constant temperature was attained by coupling 
chains of four Nos\'e-Hoover thermostats.\cite{no84,ho85} 
Another chain of four barostats was coupled to the in-plane area
of the simulation box ($xy$ plane) to yield a constant
pressure $P_{xy} = 0$.\cite{tu98,he14}
The equations of motion were integrated by using the reversible reference 
system propagator algorithm (RESPA), which permits to consider
different time steps for the integration of fast and slow
degrees of freedom.\cite{ma96}
The time step $\Delta t$ associated to the interatomic forces was 
taken as 0.5 fs, which was adequate for the atomic mass
and temperatures considered here.
The kinetic energy was calculated by employing the {\em virial}
estimator, which shows a statistical uncertainty
smaller than the {\em primitive} estimator, in particular
at high temperatures.\cite{he82,tu98}
More technical details on this type of PIMD simulations are given 
elsewhere.\cite{tu98,he06,he11}

\begin{figure}
\vspace{0.3cm}
\includegraphics[width=8.0cm]{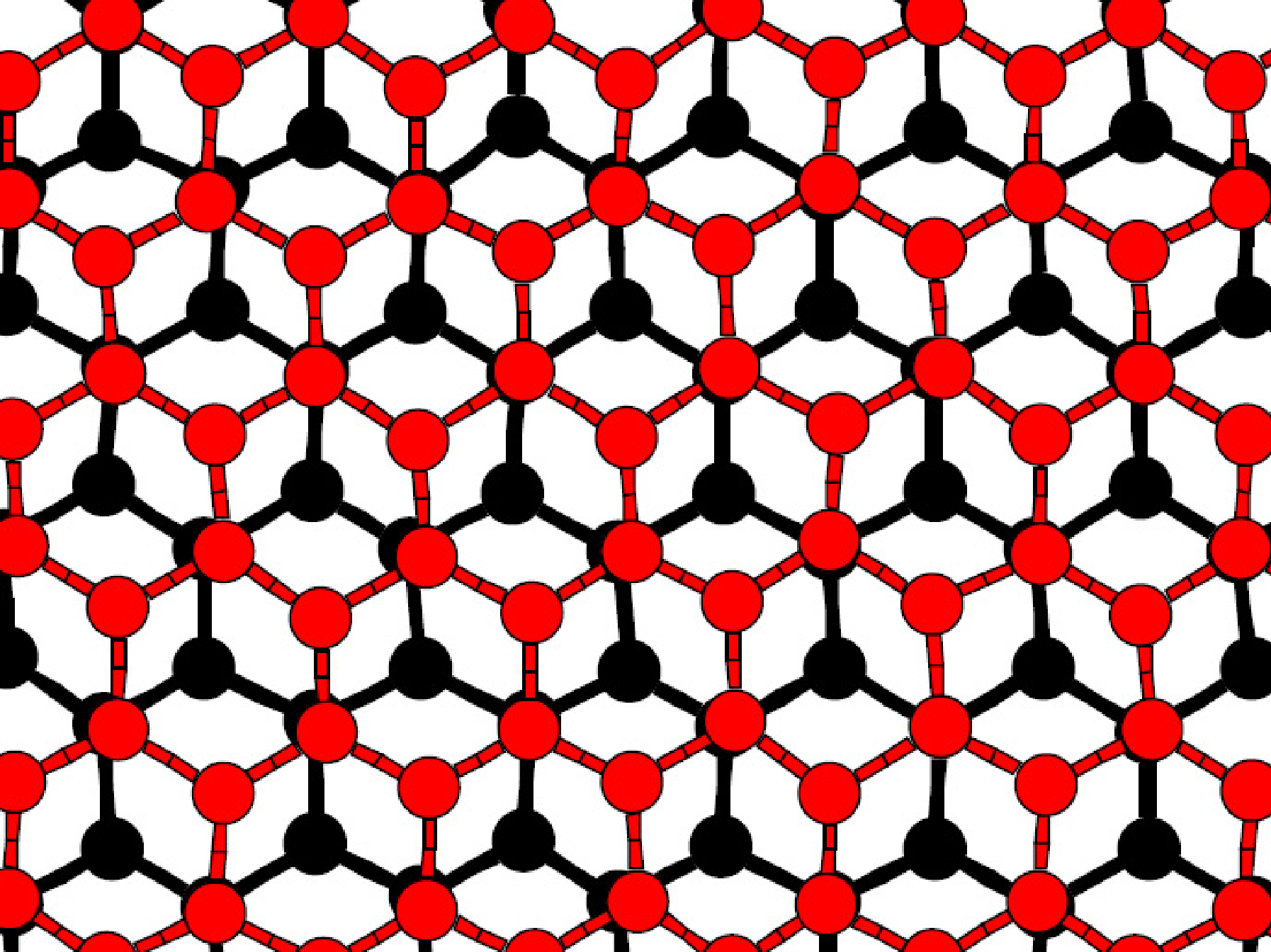}
\vspace{0.0cm}
\caption{Top view of an instantaneous configuration of bilayer
graphene at $T = 1000$~K.
Red and black circles represent carbon atoms in the upper
and lower sheets, respectively.
}
\label{f1}
\end{figure}

We have considered graphene bilayers with AB stacking in rectangular 
simulation cells including $2 N$ carbon atoms, $N$ going from 24 to 8400.
These cells had similar side lengths in the $x$ and $y$ directions
($L_x \approx L_y$), for which periodic boundary conditions were assumed.
Carbon atoms can move without restriction in the out-of-plane direction,
i.e., we have free boundary conditions in the $z$ coordinate,
reproducing a free-standing graphene bilayer.
We considered temperatures $T$ in the range from 12.5 to 1500 K.
Given a temperature, a typical simulation run consisted of
$2 \times 10^5$ PIMD equilibration steps and
$8 \times 10^6$ steps for the calculation of average variables.
The number of beads, $N_{\rm Tr}$, was taken proportional to 
$1/T$, so that $N_{\rm Tr} T$ = 6000~K, which keeps a nearly
constant accuracy for the results at different temperatures.
To assess the magnitude of nuclear quantum effects, some classical
molecular dynamics simulations of graphene bilayers have been also
carried out. This corresponds in our context 
to setting $N_{\rm Tr}$ = 1.
In Fig.~1 we present a top view of a configuration of bilayer
graphene obtained in our simulations at $T = 1000$~K.
In this picture, red and black circles represent C atoms in the upper
and lower sheets in AB stacking pattern.

For comparison with the results for graphene bilayers, we have also
performed some PIMD simulations of graphite with the interatomic
potential LCBOPII. For this 3D material we used simulation cells 
containing $4 N$ carbon atoms (four graphene sheets), and periodic
boundary conditions were assumed in the three space directions.
We used cells with $N$ = 240 and 960.

\subsection{Harmonic approximation}

To compare with the results of PIMD simulations for the specific heat 
of bilayer graphene, we will present a harmonic approximation
(HA) for the lattice vibrations. 
This approximation turns out to be rather precise at low temperature, 
but anharmonicity appears for rising temperature, so that the results
of the HA will increasingly deviate from those derived from 
the simulations. 
A basic assumption of the HA is that vibrational frequencies in the
material do not change with temperature. Then, we take in this model 
the frequencies corresponding to the minimum-energy configuration of
bilayer graphene, obtained from diagonalization of the dynamical matrix 
for the LCBOPII potential.

In a quantum HA, the vibrational energy per atom of bilayer graphene 
is given by
\begin{equation}
   E_{\rm vib}  =  \frac{1}{2N} \sum_{j,\bf k}
	\frac12 \hbar \omega_j({\bf k}) 
     \coth \left( \frac12 \beta \hbar \omega_j({\bf k}) \right)  \, ,
\label{evib}
\end{equation}
where $\beta = 1 / (k_B T)$, $k_B$ is Boltzmann's constant,
and the index $j$ ($j$ = 1, ..., 12) refers to the phonon bands:
four branches with atomic displacements along the $z$ direction
(ZA, ZO', and a two-fold degenerate ZO band), and eight branches 
with in-plane displacements (LA, TA, LO, and TO, all of them two-fold 
degenerate).\cite{ka11,ya08,si13b,ko15b}
The sum in ${\bf k}$ is extended to wavevectors
${\bf k} = (k_x, k_y)$ in the 2D hexagonal Brillouin zone,
with ${\bf k}$ points spaced by $\Delta k_x = 2 \pi / L_x$ and
$\Delta k_y = 2 \pi / L_y$.\cite{ra16}
In the following, $k$ will denote the wavenumber, 
i.e., $k = |{\bf k}|$.

The specific heat per atom, $c_v(T) = d E_{\rm vib} / d T$, 
is given in the HA by
\begin{equation}
 c_v(T) = \frac {k_B}{2N} \sum_{j,\bf k}
   \frac { \left[ \frac12 \beta \hbar \, \omega_j({\bf k}) \right]^2 }
   { \sinh^2 \left[ \frac12 \beta \hbar \, \omega_j({\bf k}) \right] } \, .
\label{cvn}
\end{equation}
Increasing the system size $N$ causes the appearance of vibrational modes
with longer wavelength $\lambda$. In fact, one has for the phonons
an effective cut-off
$\lambda_{max} \approx L$, with $L = (L_x L_y)^{1/2}$,
and the minimum wavenumber is $k_0 = 2 \pi / \lambda_{max}$, 
which means that $k_0$ scales as $N^{-1/2}$.

At low temperature, one can obtain an analytic dependence of 
the specific heat by assuming a continuous model for frequencies 
and wavenumbers, which allows to replace sums by integrals
in Eqs.~(\ref{evib}) and (\ref{cvn}). This is explained in
Sec.~VI and Appendix B.

\section{Excess area}

In our PIMD simulations in the isothermal-isobaric ensemble 
we fix the applied stress in the $(x, y)$ plane (here $P_{xy} = 0$),
as indicated in Sec.~II.A, thus allowing for changes in the in-plane 
area of the simulation cell.
Carbon atoms can freely move in the $z$ coordinate (out-of-plane 
direction), which means that at $T > 0$ the {\em real} surface 
of a graphene layer will not be planar, with an area (in 3D space) 
larger than that of the simulation cell in the $(x, y)$ plane.
The difference between the {\em real} area $A$ and {\em in-plane} 
area $A_p$ has been discussed in the literature for biological 
membranes\cite{im06,wa09,ch15} and more recently for crystalline 
membranes such as graphene.\cite{ra17}
It has been shown that values of the compressibility may
be very different when they are related to $A$ or to $A_p$.\cite{ra17}

A precise distinction between both areas is important to explain some
thermodynamic properties of 2D materials.
Thus, the area $A_p$ is the conjugate variable to the in-plane stress
$P_{xy}$ in the isothermal-isobaric ensemble used here, while the 
area $A$ is conjugate to the usually-called surface tension.\cite{sa94}
In recent years, Nicholl {\em et al.}\cite{ni15,ni17} have found that some
experimental techniques are sensitive to properties related to the
real area $A$, and other procedures can be adequate to study variables
associated to the in-plane area $A_p$.

In our PIMD simulations
we have calculated the real area $A$ of the graphene layers
by a triangulation based on the atomic positions.\cite{ra17,he19}
In the following, $A$ and $A_p = L_x L_y / N$ will refer to the real
and in-plane area per atom, respectively.
The areas $A$ and $A_p$ coincide for strictly planar graphene layers,
a condition met in the classical zero-temperature limit, 
while for $T > 0$ one has $A > A_p$.
Even for $T \to 0$, $A$ and $A_p$ are not exactly equal when
nuclear quantum effects are taken into account, due to 
zero-point motion in the out-of-plane direction.\cite{he18,he19}
For graphene monolayers and bilayers, it turns out that both areas 
present qualitatively different temperature dependencies:
the in-plane area $A_p$ displays negative thermal expansion 
in a large temperature region, while the real area $A$ does not show
that behavior.\cite{za09,he16,he19}
Moreover, $A_p$ depends on the system size, whereas
$A$ is rather insensitive to it.

\begin{figure}
\vspace{-1.0cm}
\includegraphics[width=8.0cm]{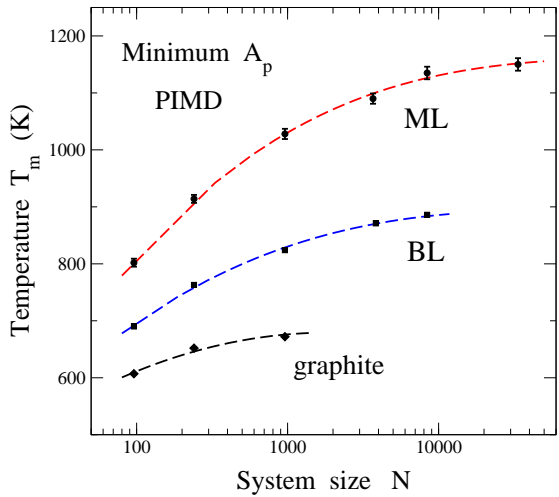}
\vspace{-0.5cm}
\caption{Temperature $T_m$ corresponding to the minimum in-plane
area $A_p$ as a function of system size.
Symbols are data points derived from PIMD simulations for
graphene monolayer (ML, circles), bilayer (BL, squares), and graphite
(diamonds). Lines are polynomial fits to the data points.
Error bars, when not displayed, are in the order or less than
the symbol size.
}
\label{f2}
\end{figure}

In the results of our PIMD simulations of graphene bilayers we observe 
that the in-plane area $A_p$ decreases as $T$ rises in the region
from $T = 0$ to temperatures of about 800 K, where it reaches a minimum,
and then it grows at higher $T$.
$A_p(T)$ presents a minimum for all considered system sizes.
This minimum becomes deeper and smoothly shifts to higher temperatures
as $N$ increases, converging to a value
$T_m = 850 (\pm 50)$ K for the largest cells considered here.
 In Fig.~2 we display the dependence of $T_m$ on system size, where
solid circles indicate results of PIMD simulations for bilayer graphene.
For comparison, we have also plotted data for monolayer graphene
(squares) and graphite (diamonds), also derived from PIMD simulations
with the LCBOPII potential model.
Dashed lines are polynomial fits to the data points.
We observe that the convergence of $T_m$ to its large-size limit
is slower for bilayer graphene than for graphite, but faster
than in the case of an isolated monolayer.
This is due to the larger out-of-plane vibrational amplitudes
in the monolayer, which are reduced in the bilayer, and are even
less for graphite.

\begin{figure}
\vspace{-0.7cm}
\includegraphics[width=8.0cm]{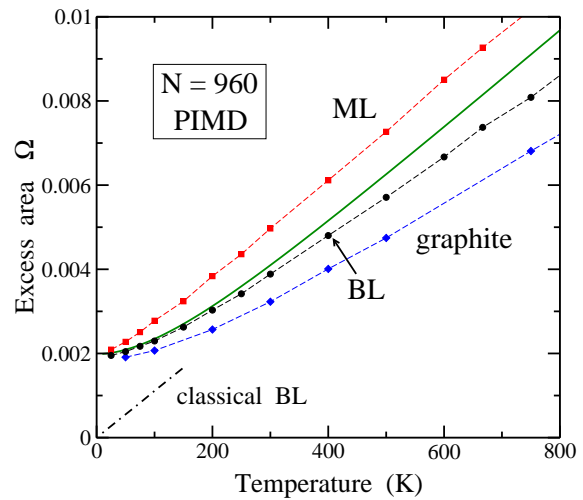}
\vspace{-0.5cm}
\caption{Temperature dependence of the dimensionless excess area,
$\Omega$, as derived from PIMD simulations for system size $N = 960$
for graphene monolayer (ML, squares), bilayer (BL, circles), and
graphite (diamonds). Dashed lines are guides to the eye.
Error bars are less than the symbol size.
The continuous line is the HA result derived from
Eq.~(\ref{omega}) in the main text.
The dashed-dotted line indicates the results of classical MD
simulations for the bilayer in the low-temperature region.
}
\label{f3}
\end{figure}

Both areas $A$ and $A_p$ derived from PIMD simulations show 
a temperature derivative which approaches zero as $T \to 0$, in 
agreement with the third law of thermodynamics.\cite{ca60,he18}
For $T \to 0$, $A$ is slightly larger than $A_p$, and the difference 
between both areas grows with temperature. 
Indeed $A_p$ is a 2D projection of the {\em real}
surface on the $(x, y)$ plane, and ripples of the actual 
surface have larger amplitudes at higher temperatures.
The difference between real and in-plane area has been called 
{\em hidden} area for graphene in Ref.~\onlinecite{ni17}, 
as well as {\em excess} area for fluid membranes.\cite{he84,fo08}
In this line, for each temperature $T$ we define the dimensionless
{\em excess} area, $\Omega$, of a graphene sheet as\cite{he84,fo08}
\begin{equation}
    \Omega = \frac{A - A_p}{A_p}  \; .
\end{equation}
In Fig.~3 we present $\Omega$ as a function of $T$ for bilayer 
graphene, as derived from our PIMD simulations (solid circles).
For comparison we also display the excess area for monolayer graphene
(squares) and graphite (diamonds). Dashed lines are guides to the eye.
The data shown here were obtained for system size $N = 960$.
In the three cases we find a low-temperature limit
$\Omega_0 = 2.0(1) \times 10^{-3}$,
i.e., $A - A_p = 5.3 \times 10^{-3}$~\AA$^2$/atom,
due to zero-point motion in the out-of-plane direction. 
As a result, the excess area grows as temperature is raised, 
in accord with an increasing amplitude of the out-of-plane 
vibrational modes.
This increase is lower for the bilayer than for the monolayer, and
it is even smaller for graphite.
Note that in a classical model $\Omega$ vanishes for $T \to 0$,
as shown in Fig.~3 for the results of classical MD simulations of 
the bilayer (dashed-dotted line).

The excess area can be calculated in a HA for the vibrational
modes, taking into account that the difference between real and
in-plane area is related to the amplitude of the modes in the
out-of-plane direction.
For a graphene sheet, the relation between 
its {\em instantaneous} real area $A_{\rm inst}$ 
and the in-plane area $A_p$ can be written in
a continuous approach as\cite{im06,wa09,ra17}
\begin{equation}
   A_{\rm inst} = 
      \int_{A_p} dx \, dy \, \sqrt{1 + |\nabla h({\bf r})|^2}  \; ,
\label{ains}
\end{equation}
where ${\bf r} \equiv (x, y)$ is the 2D position and $h({\bf r})$ is 
the distance to the mean $(x, y)$ plane of the sheet.

The difference $A_{\rm inst} - A_p$ 
can be calculated by expanding the height $h({\bf r})$ as a 
Fourier series with wavevectors ${\bf k} = (k_x, k_y)$ in the 2D hexagonal
Brillouin zone\cite{sa94,ch15,ra17} (see Appendix A).  One finds
\begin{equation}
   A = \langle A_{\rm inst} \rangle  =  
	A_p  \left[ 1 + \frac{1}{2N}  \sum_{\bf k}  k^2  
	     \langle |H({\bf k})|^2 \rangle  \right]   \; ,
\label{aains}
\end{equation}
$H({\bf k})$ being the Fourier components of $h({\bf r})$.
Then, we have for the excess area 
\begin{equation}
 \Omega =  \frac{1}{2N}
    \sum_{j,{\bf k}}  k^2  \langle |\xi_j({\bf k})|^2 \rangle  \; ,
\label{omega}
\end{equation}
with the mean-square displacements (MSDs) in a harmonic approximation
\begin{equation}
   \langle |\xi_j({\bf k})|^2 \rangle  =
       \frac{\hbar}{2 m \omega_j({\bf k})}
       \coth \left( \frac12 \beta \hbar \omega_j({\bf k}) \right)  \; .
\label{xik}
\end{equation}

For comparison with the results of our PIMD simulations, we present
in Fig.~3 the excess area $\Omega$ calculated for bilayer graphene  by
means of Eq.~(\ref{omega}) (solid line), considering the vibrational 
modes in the bands with out-of-plane displacements 
(ZA, ZO', and the two-fold degenerate ZO).
The HA yields results for the excess area close to those of the
PIMD simulations at temperatures up to 200 K.  At higher $T$, 
this approximation predicts $\Omega$ values which progressively
depart from those of the simulations, in accordance with an
increasing departure from harmonicity of the vibrational modes.

More insight into the physical meaning of the excess area as 
calculated from the MSDs $\langle |\xi_j({\bf k})|^2 \rangle$ in
Eq.~(\ref{omega}) can be obtained by looking at the classical 
(high temperature) limit in Eq.~(\ref{xik}).
In this limit, the MSDs are given by $k_B T / m \omega_j({\bf k})^2$.
This means that for low-frequency acoustic modes (LA and TA) with
$\omega \sim k$, the contribution to the sum in Eq.~(\ref{omega})
is independent of $k$. 
However, for the flexural ZA band with a negligible 
effective stress $\sigma$ ($\sigma \ll \kappa k^2$), 
one has $\omega_{\rm ZA} \approx \sqrt{\kappa} \, k^2$,
so that $k^2 \langle |\xi_j({\bf k})|^2 \rangle \sim k^{-2}$,
which makes the contribution of the flexural band the dominant
part in the sum in Eq.~(\ref{omega}).
Taking into account that the minimum wavenumber $k_0$ scales with
cell size as $k_0 \sim N^{-1/2}$ (see Sec.~II.B), its contribution
to $\Omega$ scales linearly with $N$, eventually diverging in the
thermodynamic limit. This divergence is eliminated in the presence
of an effective stress (even small) $\sigma$.
Note that in the classical limit $\Omega$ vanishes for $T \to 0$, 
at odds with the quantum result shown in Fig.~3, which
converges to a positive value $\Omega_0$ in the low-$T$ limit. 
In this limit, one has for a quantum harmonic approximation:
\begin{equation}
   \langle |\xi_j({\bf k})|^2 \rangle_0  =
       \frac{\hbar}{2 m \omega_j({\bf k})}  \; ,
\label{xik0}
\end{equation}
and the contribution to $\Omega$ of low-frequency LA and TA
modes ($\omega \sim k$) is proportional to $k$.
For the flexural ZA band we have
$\omega_{\rm ZA} \approx \sqrt{\kappa} \, k^2$,
so that  $k^2 \langle |\xi_j({\bf k})|^2$ is independent of $k$ 
and the sum in Eq.~(\ref{omega}) converges to a finite value.

\section{Thermal expansion}

In the limit $T \to 0$, the areas $A$ and $A_p$ converge to
2.6438 \AA$^2$/atom and 2.6388 \AA$^2$/atom, respectively.
For the classical minimum-energy bilayer one has a value 
of 2.6169 \AA$^2$/atom for both $A$ and $A_p$.
Then, there is a zero-point expansion of about 1\% associated
to an increase in the mean bond length, caused by
quantum zero-point vibrations (see above).
The difference between in-plane and real area (a 0.2\%)
is due to out-of-plane zero-point motion, so that even at
$T = 0$ the graphene layers are not totally planar, 
as indicated above.

\begin{figure}
\vspace{-0.7cm}
\includegraphics[width=8.0cm]{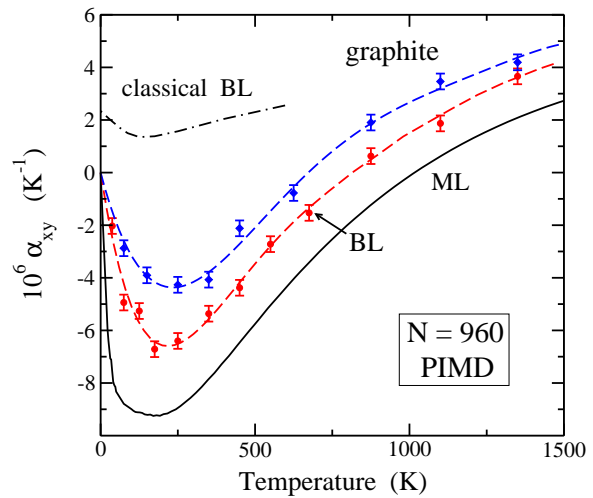}
\vspace{-0.5cm}
\caption{In-plane thermal expansion coefficients $\alpha_{xy}$
vs temperature, as derived from PIMD simulations for cell size
$N = 960$. Symbols are data points obtained from numerical
derivatives of $A_p$ for graphene bilayer (BL, circles) and graphite
(diamonds). Dashed lines are polynomial fits to the data points.
The solid line is a fit to earlier results for a graphene
monolayer (ML).\cite{he18}
The dashed-dotted line indicates the classical result for the bilayer
for temperature up to 600~K.
}
\label{f4}
\end{figure}

Associated to the area $A_p$, we define an in-plane thermal 
expansion coefficient (TEC) as
\begin{equation}
 \alpha_{xy} = \frac{1}{A_p}
      \left( \frac{\partial A_p}{\partial T} \right)_{P_{xy}}   \, .
\end{equation}
In Fig.~4 we show $\alpha_{xy}$ derived from our PIMD simulations 
for bilayer graphene (solid circles).
These data points were obtained from a numerical derivative
of the area $A_p$ found in the simulations.
For comparison we also display results for graphite (diamonds).
In both cases, the dashed lines represent polynomial fits to the
data points. The solid line indicates a fit to results of PIMD
simulations of monolayer graphene presented in Ref.~\onlinecite{he18}.
All these results correspond to a system size $N = 960$.
In the three cases, $\alpha_{xy}$ vanishes in the low-temperature limit,
in line with the third law of Thermodynamics.

The general trend of $\alpha_{xy}$ vs $T$ is similar in the three
cases shown in Fig.~4: at low temperatures $\alpha_{xy}$ decreases for 
rising $T$ and reaches a minimum at a temperature $T_m'$.
The main difference between them appears
in the magnitude of the minimum of the curves.
Moreover, $T_m'$ increases from a value of 180 K for monolayer graphene
to 235 K for graphite.
At low temperatures, $\alpha_{xy}$ decreases fast for increasing $T$, 
and for the bilayer it attains a minimum amounting to
$-6.6 \times 10^{-6}$ K$^{-1}$ at $T_m' \approx 220$~K.
At higher $T$, $\alpha_{xy}$ approaches zero and becomes
positive at $T_m$ = 820 K (where $A_p$ takes its minimum value,
see Fig.~2).
At $T > 500$~K the three materials present almost the same
dependence of $\alpha_{xy}$ on $T$, apart from rigid shifts of the
corresponding curves.
Our results for $\alpha_{xy}$ presented in Fig.~4 are qualitatively
similar to those derived earlier for monolayer graphene
from other theoretical techniques\cite{ji09,si14} and
experimental methods.\cite{yo11,ba09}

For graphite, experimental data of the area TEC $\alpha_{xy}$
display a minimum at a temperature between 200 and 300 K, 
similar to that derived from our 
simulations.\cite{ke64,mo72,ma18} Various data present a minimum 
of $\approx -3 \times 10^{-6}$~K$^{-1}$, 
somewhat smaller than our result for graphite shown in Fig.~4.

The behavior of the in-plane TEC as a function of temperature 
can be understood as due to two opposing contributions.
First, there appears  a trend of the C--C distance to grow
as $T$ is raised, thus favoring an increase in $A_p$.
Second, bending of the graphene sheets
causes a reduction of its projection on the $(x, y)$ plane,
i.e. the in-pane area $A_p$.
At low $T$, the rise of the in-plane area caused by the first 
contribution (bond expansion) is overshadowed 
by the second one (bending), and $d A_p / d T < 0$.
At high $T$, the increase in C--C distance dominates the reduction
in $A_p$ due to out-of-plane atomic displacements,
so that one has $d A_p / d T > 0$.
The increase in $T_m$ for rising system size shown in Fig.~2 
is a consequence of the growth of the out-of-plane bending
of the graphene sheets for larger $N$.

One can equally define a TEC 
$\alpha = (\partial A/ \partial T) / A$ 
for the real area of the graphene sheets.
The real area behaves as a function of $T$ in an analogous way
to the crystal volume of most 3D solids,\cite{as76}
i.e., it increases at all finite temperatures.
The area TEC $\alpha$ is insensitive to the system size,\cite{he19} 
and coincides within error bars for monolayer, bilayer graphene, and
graphite.

In the low-temperature limit our PIMD simulations yield for bilayer
graphene an interlayer spacing, $c$, of 3.3520~\AA, to be
compared with that corresponding to the classical minimum:
$c_0$ = 3.3372~\AA\ (planar graphene sheets in AB stacking).
This means a zero-point expansion of $1.5 \times 10^{-2}$ \AA,
i.e., the mean spacing between layers increases by a 0.5\% with
respect to the classical prediction.
At $T$ = 300 K, PIMD simulations give $c$ = 3.3758~\AA, and
the difference between classical and quantum results is about
five times less than in the low-temperature limit.\cite{he19}

\begin{figure}
\vspace{-0.7cm}
\includegraphics[width=8.0cm]{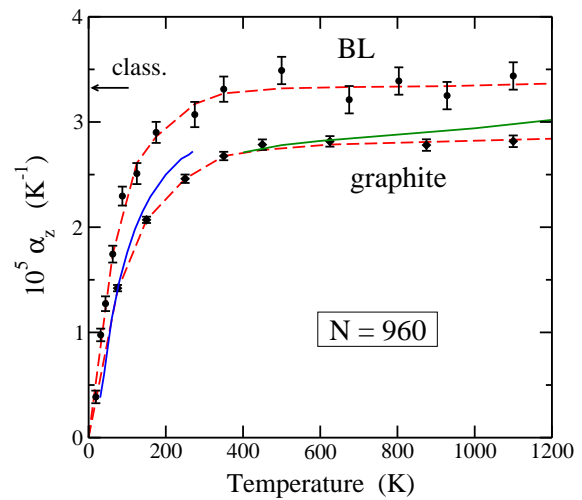}
\vspace{-0.5cm}
\caption{Thermal expansion coefficient $\alpha_z$ in the
out-of-plane direction vs the temperature, as derived from PIMD
simulations for $N = 960$.
Symbols are data points obtained from temperature derivatives of
of the interlayer spacing for bilayer graphene (BL, circles)
and graphite (diamonds).
Dashed lines are guides to the eye.
The blue solid line represents data obtained for graphite by
Bailey and Yates\cite{ba70} from interferometric measurements
at $T < 300$~K. The green solid line is a fit to experimental data
of graphite for $T > 300$~K, presented by
Marsden {\em et al.}\cite{ma18}
A horizontal arrow indicates the low-temperature limit of
the classical simulations for the bilayer (labeled as ``class'').
}
\label{f5}
\end{figure}

From the mean interlayer spacing we define the out-of-plane 
TEC $\alpha_z$ as
\begin{equation}
 \alpha_z = \frac{1}{c}
      \left( \frac{\partial c}{\partial T} \right)_{P_{xy}}   \, .
\end{equation}
This TEC has been usually called $\alpha_c$ in the graphite 
literature, but we will call it here $\alpha_z$ for consistency
of our notation.
In Fig.~5 we show results for $\alpha_z$ derived from our
PIMD simulations for bilayer graphene (solid circles) and graphite
(diamonds). Dashed lines are guides to the eye.
$\alpha_z$ turns out to be higher for the bilayer than for graphite
at all finite temperatures, since the graphene layers are more
free to move in the out-of-plane direction in the bilayer,
as compared to graphite.
In both cases one observes a fast increase in $\alpha_z$ up to
about 200 K, which becomes rather slow for $T > 400$~K.

A blue solid line in Fig.~5 represents $\alpha_z$ data obtained for 
pyrolytic graphite by Bailey and Yates\cite{ba70} from interferometric 
measurements at low temperatures.
A green solid line represents a fit to experimental data for 
graphite at $T > 300$~K.\cite{ma18}
Both lines fitted to experimental results do not match well one with
the other, mainly due to data dispersion in different source 
references. At high temperature, one observes that $\alpha_z$ 
derived from our PIMD
simulations increases slower than the experimental data.

To end this section, we comment on the fact that classical atomistic
simulations cannot give reliable results for several properties
of graphene (condensed matter in general) at temperatures below
the Debye temperature of the material, $\Theta_D$.\cite{ki66,as76}
This is the case of thermal expansion coefficients, which have to
vanish in the low-temperature limit, according to the third law
of Thermodynamics.\cite{ca60} 
In our case of graphene bilayers, classical simulations 
yield unphysical finite (positive) values for $\alpha_{xy}$ and 
$\alpha_z$ when $T \to 0$, as indicated in Figs.~4 and 5
by a dashed-dotted line and an arrow, respectively. This failure of
classical simulations is the same as that known for solids when
atomic vibrations are described by classical models,\cite{ki66,as76}
and has been observed earlier for monolayer graphene.\cite{he16}

In classical simulations, the vibrational states display a (nonrealistic) 
continuous energy distribution that causes physical anomalies at low 
temperatures.  This is related to the quantization of vibrational states, 
which is adequately described by path-integral simulations. 
Since the actual values of $\alpha_{xy}$ and $\alpha_z$ are given by the 
relative population of the excited vibrational states, both variables
converge to zero for $T \to 0$ due to the presence of the energy gap for 
the vibrational modes.
The failure of classical simulations is remedied at relatively high $T$ 
(the scale is set by $\Theta_D$), when excited states
are appreciably populated.  
We finally note that simultaneous anomalies at low temperature in 
thermal expansion coefficients and the specific heat $c_p$ derived from 
classical simulations are expected from the thermodynamic relations 
between these variables.\cite{ca60,as76,he18}

\section{Compressibility}

\subsection{In-plane compressibility}

PIMD simulations allow one to obtain insight into the elastic
properties of materials under different conditions, i.e., 
various kinds of external stresses such as hydrostatic or
uniaxial.
For a two-dimensional material, we understand a {\em hydrostatic}
stress in a similar way to three-dimensional materials,
but applied in a plane (with units of force per unit length).
In the language of elasticity this means in our case
$\sigma_{xx} = \sigma_{yy} = P_{xy}$ (see Ref.~\onlinecite{be96b}).
Then, we define the in-plane isothermal compressibility 
per layer as
\begin{equation}
   \chi_{xy} = - \frac{n}{A_p}
       \left( \frac{\partial A_p}{\partial P_{xy}} \right)_T   \, .
\label{chip1}
\end{equation}
where $n$ is the number of layers, i.e., $n = 1$ for the monolayer
and $n = 2 $ for bilayer graphene.
In this equation, the variables on the r.h.s. correspond to
in-plane quantities, since the pressure $P_{xy}$ in the
isothermal-isobaric ensemble employed here is the conjugate variable 
to the in-plane area $A_p$.
Note that the normalizing factor $n$ appears in the numerator 
in Eq.~(\ref{chip1}), because the inverse of the compressibility
(the 2D modulus of hydrostatic compression\cite{be96b}) 
is an extensive magnitude proportional to the number of layers.

\begin{figure}
\vspace{-0.7cm}
\includegraphics[width=8.0cm]{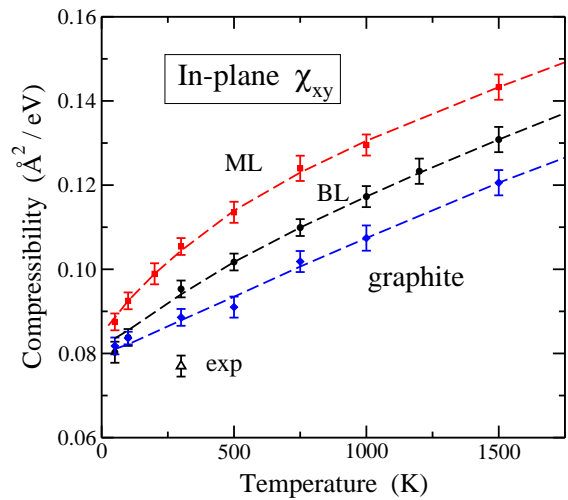}
\vspace{-0.5cm}
\caption{
Temperature dependence of the compressibility $\chi_{xy}$, as derived
from PIMD simulations for monolayer (ML, squares), bilayer graphene
(BL, circles) and graphite (diamonds).
Lines are guides to the eye.
An open triangle (labeled "exp") indicates the result derived from
experimental data for graphite.\cite{bl70}
}
\label{f6}
\end{figure}

An alternative way to calculate the compressibility $\chi_{xy}$
is based on the fluctuation formula\cite{la80,ra17,he18b}
\begin{equation}
   \chi_{xy} = \frac{n N \Delta_p^2}{k_B T A_p}   \; ,
\label{chip2}
\end{equation}
where $\Delta_p^2$ are the mean-square fluctuations of the
area $A_p$, which in our case are obtained from PIMD simulations
at $P_{xy} = 0$.
In our context, this formula turns out to be more convenient than
calculating $(\partial A_p / \partial P_{xy})_T$, because obtaining 
this derivative by numerical procedures involves additional 
simulations at nonzero stresses.
For some selected temperatures we have checked that both methods yield 
the same results for $\chi_{xy}$, inside the statistical error bars.

In Fig.~6 we show the temperature dependence of the compressibility
$\chi_{xy}$ of bilayer graphene (solid circles), as derived 
from our PIMD simulations by using Eq.~(\ref{chip2}).
For comparison we also present results for monolayer graphene
(squares), as well as for graphite (diamonds).
At low $T$, we find in the three cases compressibility values
close to $\chi_{xy}$ = 0.08 \AA$^2$/eV, and the difference between
them becomes larger as temperature is raised.
$\chi_{xy}$ for bilayer graphene is intermediate between those of
monolayer graphene and graphite. 
Interactions between layers cause a reduction in the out-of-plane
vibrational amplitudes of the carbon atoms, so that the
layers effectively become ``harder'', i.e., the in-plane
compressibility decreases.
Something similar is observed for the out-of-plane compressibility
$\chi_z$, as shown below.

As noted above, the inverse of $\chi_{xy}$, $B_{xy} = 1 / \chi_{xy}$, 
is the 2D modulus of hydrostatic compression,\cite{be96b} with 
units of eV/\AA$^2$ or N/m.
For graphene, with in-plane hexagonal symmetry, $B_{xy}$ may be
written as a function of the elastic constants of 
the material as
\begin{equation}
	B_{xy} = \frac{1}{2} (c_{11} + c_{12})   \; ,
\end{equation}
These elastic constants are related with
the Lam\'e parameters, $\mu$ and $\lambda$ by 
$c_{11} = \lambda + 2 \mu$ and $c_{12} = \lambda$, so that
$B_{xy} = \lambda + \mu$.\cite{be96b}

In the case of graphite, we can make connection of the results
obtained here for $\chi_{xy}$ with material properties derived from
experiment. For this purpose, we can convert the elastic
constants of graphite $C_{11}$ and $C_{12}$ (units of force per
square length) into in-plane elastic constants $c_{ij}$ as
$c_{ij} = c \, C_{ij}$, using the mean interlayer distance $c$.
Then, for graphite we take $C_{11} = 1060 \pm 20$ GPa, 
$C_{12} = 180 \pm 20$ GPa,\cite{bl70} and $c$ = 3.3538~\AA,\cite{ba55}
and find $\chi_{xy} = 1 / B_{xy}$ = 0.077(2) \AA$^2$/eV. For comparison
with the results of our simulations, this data point is shown
in Fig.~6 as an open triangle at 300 K.
The result of our simulations for graphite at $T$ = 300 K is 
somewhat higher than that derived from experimental data.

\subsection{Out-of-plane compressibility}

We now turn to the compressibility $\chi_z$ of bilayer graphene 
in the out-of-plane direction. Similarly to the in-plane
compressibility $\chi_{xy}$, $\chi_z$ can be calculated from 
the interlayer spacing and its fluctuations along a simulation 
run at a given temperature.
The isothermal compressibility in the $z$ direction is
defined as
\begin{equation}
  \chi_z = - \frac{1}{V}
            \frac{\partial V}{\partial P_z}  \, ,
\label{chiz}
\end{equation}
where $V = c L_x L_y$ and $P_z$ is a uniaxial stress in the
out-of-plane direction.
The compressibility $\chi_z$ of bilayer graphene at a temperature
$T$ may be calculated from PIMD simulations with $P_z = 0$ by 
employing the fluctuation formula\cite{la80,he08}
\begin{equation}
    \chi_z = \frac{\Delta_V^2}{k_B T V}  \, ,
\label{chiz2}
\end{equation}
where the volume mean-square fluctuations associated to changes in the
interlayer distance $c$ are given by
$\Delta_V^2 = L_x^2 L_y^2 \Delta_c^2$.
Then, we obtain $\chi_z$ by using the expression
\begin{equation}
   \chi_z = \frac{L_x L_y}{k_B T} \frac{\Delta_c^2}{c}   \, .
\label{chiz3}
\end{equation}
Note that in this expression $L_x$, $L_y$, and $c$ indicate mean
values of these variables along a simulation run at temperature $T$.

\begin{figure}
\vspace{-0.7cm}
\includegraphics[width=8.0cm]{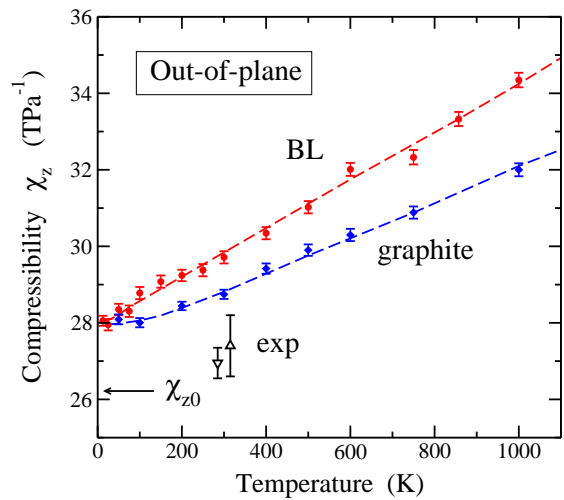}
\vspace{-0.5cm}
\caption{
Temperature dependence of the compressibility $\chi_z$, as derived
from PIMD simulations for bilayer graphene (BL, circles) and
graphite (diamonds).  Lines are guides to the eye.
Open triangles indicate results derived from experimental data
of graphite at room temperature:
triangle up from Ref.~\onlinecite{bl70} and
triangle down from Ref.~\onlinecite{ni72}.
A horizontal arrow shows the classical zero-temperature
limit $\chi_{z0}$.
}
\label{f7}
\end{figure}

The temperature dependence of $\chi_z$ is shown in Fig.~7.
Solid circles are data points obtained from our PIMD simulations
for bilayer graphene.
Besides, we display in Fig.~7 data for the compressibility of graphite,
derived also from PIMD simulations, using Eq.~(\ref{chiz3}).
Both sets of results converge at low temperature to the same value
of the compressibility (within error bars): 
$\chi_z = 2.79(2) \times 10^{-2}$ GPa$^{-1}$, because 
the MSDs $\Delta_c^2$ are found to be 
nearly identical for bilayer graphene and graphite.
For higher $T$, $\Delta_c^2$ is smaller for graphite, and 
therefore its compressibility $\chi_z$ is lower than that of 
bilayer graphene.

We note that the classical compressibility $\chi_{z0}$ for $T \to 0$
can be calculated from the dependence of the system energy on the
interlayer spacing $c$ close to the minimum-energy value $c_0$. 
This yields $\chi_{z0} = 2.63 \times 10^{-12}$ cm$^2$ dyn$^{-1}$ or
0.0263 GPa$^{-1}$ (see Ref.~\onlinecite{he19}), a value
indicated in Fig.~7 by a horizontal arrow.
This means an appreciable increase of a 6\% in the low-temperature 
quantum value of $\chi_z$ with respect to the classical limit.

The compressibility $\chi_z$ coincides in the case of graphite 
with the elastic compliance constant $S_{33}$ of this material, 
since this constant connects stress and strain 
in the $z$ direction.\cite{ma18}
In Fig.~7 we show $S_{33}$ obtained for pyrolytic graphite 
from neutron diffraction data combined with a force model\cite{ni72}
(triangle down), and from ultrasonic test methods\cite{bl70} 
(triangle up). These data were obtained at room temperature 
and are horizontally moved around 300 K in Fig.~7 for the sake 
of clarity. Note that for graphite $S_{33}$ is related to the
elastic constant $C_{33}$ as $S_{33} C_{33} \approx 1$, and the
difference between $S_{33}$ and $C_{33}^{-1}$ is less than the
error bars of the experimental data.\cite{ni72,bl70}
Our results overestimate the compressibility $\chi_z$ of graphite
by nearly a 5\% with respect to those data derived from experiments
at room temperature.

Komatsu\cite{ko64} found at low-temperature ($T \approx 2$~K) 
a value of the elastic constant $C_{33} = 35.6$~GPa from 
specific-heat measurements of natural and pile graphite, which 
translates to $\chi_z$ = 0.0282 GPa$^{-1}$.
This value (not shown in Fig.~7) is close to our results 
for graphite, but there is no available error bar for it.

\section{Specific heat}

The calculation of low-temperature specific heats of materials 
by means of path-integral simulations is not straightforward in general.
Even obtaining the Debye law $c_p \sim T^3$ for 3D solids has been 
a challenge for PIMD, because of the effective low-frequency 
cut-off associated to the finite size of the simulation 
cells.\cite{no96,ra06}
This situation is improved in simulations of 2D materials such as
graphene, mainly for two reasons.
First, the length of the cell sides scales as $L \sim N^{1/d}$
($d$, dimension of the space), and
the minimum wavenumber $k_0$ available in the simulation 
scales as $k_0 \sim N^{-1/d}$.
Thus, for increasing number of atoms, $k_0$ decreases faster 
for $d = 2$ than for $d = 3$.
This means that the low-frequency region is described better for
2D materials, and therefore also the low-temperature region.
Second, the internal energy for graphene rises at low temperature 
as $T^2$ (i.e., $c_p \sim T$), which is a fast increase at low 
temperature, when compared with the typical expectancy ($E \sim T^4$) 
for the phonon contribution in 3D materials ($c_p \sim T^3$).

\begin{figure}
\vspace{-1.0cm}
\includegraphics[width=8.0cm]{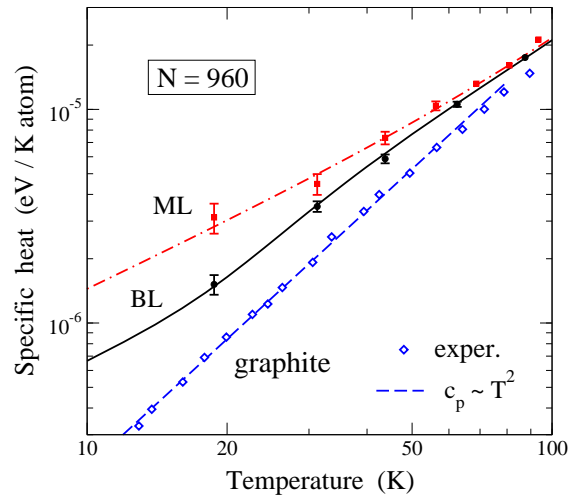}
\vspace{-0.5cm}
\caption{Specific heat of graphene as a function of temperature.
Solid symbols represent results for $c_p$ derived from PIMD simulations
for $N = 960$: squares for graphene monolayer and circles for the bilayer.
The solid line is $c_v$ obtained from the 12 phonon bands of bilayer
graphene, corresponding to the LCBOPII potential in a harmonic
approximation. The dashed-dotted line indicates $c_v$ for monolayer
graphene in the HA.\cite{he18}
Open diamonds represent experimental data for graphite obtained by
Desorbo and Tyler.\cite{de53}
The dashed line shows the dependence $c_p \propto T^2$.
}
\label{f8}
\end{figure}

The specific heat of graphene is controlled by the vibrational
contribution, the electronic part $c_p^{\rm el}$ being negligible 
with respect to the former.
In fact, $c_p^{\rm el}$ has been estimated in various works, and
it results to be between three and four orders of magnitude less than
the vibrational part.\cite{be96,ni03,fo13}

In Fig.~8 we present the temperature dependence of the specific
heat in the low-temperature region in a logarithmic plot.
Solid symbols are results for $c_p$ obtained from PIMD simulations for
$N$ = 960: circles for a graphene bilayer and squares for 
a monolayer.  They were obtained from a
numerical derivative of the internal energy $E(T)$.
The solid line represents $c_v$ for the bilayer, derived from 
the harmonic approximation given by Eq.~(\ref{cvn}) for the same 
cell size.
The dashed-dotted line indicates the HA for the monolayer, obtained
by adding the contributions of the six phonon bands appearing in
this case.\cite{he18}

We first observe in Fig.~8 a good agreement between the results 
of the HA and those derived from PIMD simulations at $T < 100$~K 
in both cases, monolayer and bilayer graphene.
Results of the simulations are close to the HA up to about 300~K, and
at higher $T$ they gradually depart from the solid line,
in a temperature region where anharmonic effects are expected to 
be observable.
For monolayer graphene, one observes a linear dependence
of the specific heat for $T \lesssim 40$~K 
(slope unity in the logarithmic plot),
given by $c_p \approx C T$ with $C = 1.4 \times 10^{-7}$ eV K$^{-2}$
For the bilayer, a similar trend with a linear dependence of
$c_p$ appears also at low $T$ with a constant
$C = 5.0 \times 10^{-8}$ eV K$^{-2}$, but for $T \gtrsim 15$~K
the temperature dependence becomes superlinear.
This trend is explained below.

For comparison with the results of our simulations, we also show 
in Fig.~8 experimental data for $c_p$ of graphite, obtained by 
Desorbo and Tyler from calorimetric measurements\cite{de53}
(open diamonds).
The specific heat of graphite has been thoroughly analyzed
in a wide range of temperatures.\cite{ko51,kr53,kl53,ni72}
For this 3D material, $c_p$ increases as $T^3$ for $T < 10$~K 
(a region not reached in our simulations and not presented in Fig.~8). 
For $T$ between 10 and 100~K, $c_p$ rises as $T^2$, a typical 
dependence in strongly anisotropic solids.\cite{ni72,ni03}
The most important difference between graphite and graphene 
(monolayer and bilayer) in this temperature range  
consists in the dominant contribution to $c_p$ coming from 
phonons with linear dispersion relation 
($\omega \sim k$) for small $k$ in graphite.
At room temperature ($T = 300$~K) the measured specific heat of 
graphite equals $8.90 \times 10^{-5}$ eV/(K atom), 
or 8.59 J/(K mol),\cite{de53}
to be compared with the result of our PIMD simulations for
bilayer graphene, $c_p = 9.2(\pm 0.1) \times 10^{-5}$ eV/(K atom),
and for a monolayer, 
$c_p = 9.4(\pm 0.1) \times 10^{-5}$ eV/(K atom).\cite{he18}

One can also calculate the specific heat $c_v$ from constant-$A_p$
simulations, analogous to $NVT$ simulations in 3D materials.
From thermodynamic considerations, one should have
$c_v \leq c_p$ at any temperature, but the difference
between them for bilayer graphene is smaller than the statistical
error bar of our numerical results, and they are indistinguishable
in the results derived from PIMD simulations.\cite{he18}

The difference between $c_p$ and $c_v$ can be obtained from
the formula\cite{he18}  
\begin{equation}
    c_p - c_v = \frac{n T \alpha_{xy}^2 A_p}{\chi_{xy}}  \,
\label{cpcv}
\end{equation}
which is similar to the well-known thermodynamic expression for
this difference of specific heats in 3D systems.\cite{ca60,la80}
The variables present on the r.h.s. of Eq.~(\ref{cpcv})
refer to in-plane properties, since the pressure appearing in our
isothermal-isobaric ensemble is the conjugate variable of 
the in-plane area $A_p$.

For thermodynamic consistency one needs $c_p \geq c_v$, in accord
with Eq.~(\ref{cpcv}), and we have $c_p - c_v = 0$ whenever 
$\alpha_{xy}$ vanishes. This happens for bilayer graphene at 
$T_m \approx 850$~K, as shown in Sec.~III (apart from the trivial
coincidence $c_p = c_v = 0$ at $T = 0$). In the interval from
$T = 0$ to 850~K, the maximum difference is reached at
$T \approx 200$~K, close to the maximum of $|\alpha_{xy}|$, where 
we find using Eq.~(\ref{cpcv}):  
$c_p - c_v = 5.6 \times 10^{-7}$ eV/(K atom).
For $T < 50$~K, we have
$c_p - c_v < 4 \times 10^{-8}$ eV/(K atom), less than the
statistical error bars of the results for $c_p$ derived from
our PIMD simulations.

\begin{figure}
\vspace{-0.7cm}
\includegraphics[width=8.0cm]{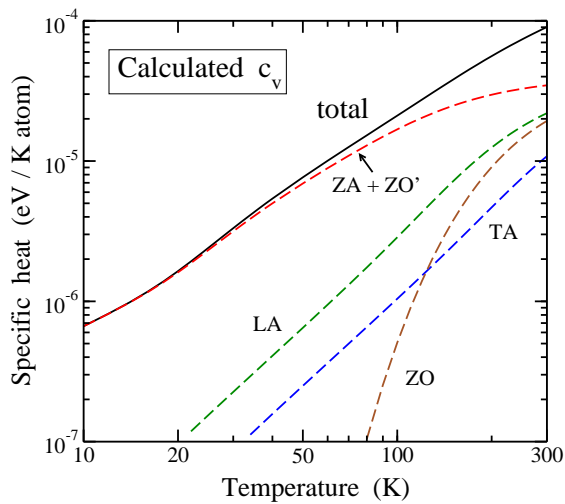}
\vspace{-0.5cm}
\caption{Contributions $c_v^j$ of the phonon bands to the
specific heat of bilayer graphene as a function of temperature.
}
\label{f9}
\end{figure}

The low-temperature behavior of the specific heat can be
analyzed by considering a continuous model for frequencies and
wavenumbers, as in the well-known Debye model for
solids\cite{as76} (see Appendix B).
At low $T$, $c_v$ is controlled by
the input of acoustic modes with small $k$.
In the case of graphene, these are TA and LA modes
with $\omega_j \propto k$ and ZA modes with $\omega_j \propto k^2$.
Note that an effective stress $\sigma$ introduces
a linear contribution for ZA modes of small $k$, but this
will be negligible for the temperatures considered here
and vanishing external in-plane stress.
For bilayer graphene, it is appreciable at $T > 15$~K the role of
the layer-breathing ZO' band, which is nearly flat close to 
the $\Gamma$ point ($k = 0$), with a frequency 
$\omega_0 = 92$~cm$^{-1}$.\cite{ya08,he19}

To understand the behavior of the specific heat of bilayer graphene
in the temperature region displayed in Fig.~8, we discuss the
contributions of the different phonon branches in the HA.
In Fig.~9 we present as dashed lines these contributions as a function
of $T$. At temperatures lower than 10~K, the specific heat is controlled 
by the flexural ZA modes with $\omega(k) \sim k^2$ and 
out-of-plane displacements. This gives $c_v^j \sim T$, 
as in the case of monolayer graphene (see Appendix B).
For the bilayer, however, the input of the ZO' band is
relevant for $T \gtrsim 15$~K, and $c_v$ appreciably departs from
linearity. The contributions of the acoustic LA and TA branches
($\omega \sim k$) appear in Fig.~9 for $T > 20$~K with 
a slope of two, i.e., $c_v^j \sim T^2$.
The input of the optical ZO band increases exponentially
at low $T$ and becomes observable at temperatures in the order of
100~K. The other optical bands (LO and TO), with higher frequencies,
are important for the specific heat at $T > 300$~K.

The specific heat of graphene bilayer is lower than that of
the monolayer due to the relative contributions of acoustic
phonons at low $T$. In particular, the shape of the flexural 
ZA band is nearly the same in both cases, and it is not degenerate.
This means that it contributes less in bilayer graphene
(12 bands) than in the monolayer (6 bands).

We finally note that several atomistic simulations of graphene 
monolayers and bilayers have been carried out
in the past using classical Monte Carlo and molecular dynamics
simulations. These are well-established methods to study structural,
dynamical, and thermodynamic properties in condensed matter, but
some of these properties may be far from the corresponding
{\em real} values at temperatures lower than the Debye temperature
$\Theta_D$ of the considered material,\cite{as76} as indicated
for the thermal expansion in Sec.~IV.
Thus, values presented in the literature for the specific heat
of graphene monolayers\cite{la12} 
and bilayers,\cite{za10b} derived from classical
simulations, are close to the Dulong-Petit specific heat,
i.e., $c_v^{\rm cl} = 3 k_B$. This means in our units
$c_v^{\rm cl} = 2.6 \times 10^{-4}$ eV/(K atom), which turns out
to be about three times larger than the value obtained from
our quantum PIMD simulations at 300 K. 
The difference between classical and quantum results increases as
temperature is lowered, and at $T$ = 20 K the classical value is 
two orders of magnitude larger than the quantum result. 
Even at $T$ = 1000 K 
the quantum data are still appreciably lower than the
classical limit.

\section{Summary}

PIMD simulations have revealed as a suitable tool to study
thermodynamic properties of graphene bilayers.
In this paper we have presented results
obtained in the isothermal-isobaric ensemble in a wide range of
temperatures and zero external stress.
We have concentrated on physical properties as the excess area,
thermal expansion, in-plane and out-of-plane compressibility,
and specific heat.
Explicit consideration of the quantum character of atomic nuclei
is crucial for a realistic description of these crystalline
membranes, even for $T$ higher than room temperature.
This is particularly important for the heat capacity and
compressibility.

A thermal contraction of the in-plane area $A_p$ appears in bilayer
graphene in a similar way to an isolated monolayer, although
this contraction is less important in the former case.
This is due to a reduction of out-of-plane vibrational
amplitudes of the C atoms in the bilayer, associated to
interlayer interactions.
We find a negative $\alpha_{xy}$ for $T \lesssim$ 800~K,
and it becomes positive at higher temperature.
The difference $A - A_p$ between the real area $A$ and the in-plane
area $A_p$ grows as temperature rises and deviations from planarity
of the graphene sheets become more appreciable.
This has been quantified by the
dimensionless excess area $\Omega$, which converges to a value
$\Omega_0 = 2 \times 10^{-3}$ for $T \to 0$,
due to quantum zero-point motion.

The in-plane $\chi_{xy}$ and out-of-plane $\chi_z$ compressibilities
of graphene bilayers have been obtained from the fluctuations of
the in-plane area and the interlayer distance, respectively.
This procedure accurately yields the increase in $\chi_{xy}$
and $\chi_z$ as $T$ is raised.

Comparison of our simulation results with those yielded by a
HA for the vibrational modes has allowed us
to assess the effects of anharmonicity in finite-temperature
properties of graphene bilayers.
Such anharmonicity clearly shows up at temperatures higher
than 200 K, as shown in Fig.~3 for the excess area.
At lower temperatures, however, thermal properties of
the graphene bilayers considered here are well described
by the HA, using the vibrational frequencies obtained for
the classical equilibrium geometry at $T = 0$.

At the lowest temperatures studied here ($T >$ 10~K), the
HA predicts a linear dependence of the specific heat
$c_v = C T$, with $C = 5.0 \times 10^{-8}$ eV K$^{-2}$, and
for $T \gtrsim 15$~K the temperature dependence becomes
superlinear, in agreement with the results of our PIMD
simulations. This trend is different than that corresponding
to monolayer graphene, due to the contribution of the
layer-breathing ZO' band in the case of the bilayer.

PIMD simulations similar to those presented here can provide
insight about the thermal properties of free-standing graphene
multilayers under tensile and compressive stress.
This would give information on the relative stability of these
multilayers in a stress-temperature phase diagram.

\begin{acknowledgments}
The authors acknowledge the help of J. H. Los in the implementation 
of the LCBOPII potential.  This work was supported by
Ministerio de Ciencia, Innovaci\'on y Universidades (Spain) through
Grants FIS2015-64222-C2 and PGC2018-096955-B-C44.
\end{acknowledgments}

%  -----------------------------------------------------------------

\appendix

\section{Calculation of the excess area}

In the continuum limit, the instantaneous real area $A_{\rm inst}$ 
of a graphene sheet is given by\cite{im06,wa09,ra17}
\begin{equation}
 A_{\rm inst} = \int_{A_p} dx \, dy \, \sqrt{1+|\nabla h({\bf r})|^2}  \; ,
\label{aap2}
\end{equation}
where ${\bf r} \equiv (x, y)$ indicates the 2D position and
$h({\bf r})$ is the height of the surface, i.e. the distance to
the mean $(x, y)$ plane of the sheet.
For small $|\nabla h({\bf r})|$ (in fact for 
$(\partial h / \partial x)^2 + (\partial h / \partial y)^2 \ll 1$,
which is the case here), one has
\begin{equation}
  A_{\rm inst} \approx \int_{A_p} dx \, dy \, 
	\left[ 1 + \frac12 |\nabla h({\bf r})|^2  \right]    \; .
\label{aap3}
\end{equation}

We now write the out-of-plane displacement $h({\bf r})$ as a Fourier
series
\begin{equation}
    h({\bf r}) = \frac{1}{\sqrt{N}} 
        \sum_{\bf k} {\rm e}^{i {\bf k \cdot r}} H({\bf k})
\label{hxy}
\end{equation}
with wavevectors ${\bf k} = (k_x, k_y)$ in the 2D hexagonal 
Brillouin zone, i.e., $k_x = 2 \pi n_x/ L_x$ and
$k_y = 2 \pi n_y/ L_y$ with integers $n_x$ and $n_y$.\cite{ra16}  
The Fourier components are given by
\begin{equation}
   H({\bf k}) = \frac{\sqrt{N}}{A_p}  \int_{A_p} dx \, dy \, 
	 {\rm e}^{- i {\bf k \cdot r}}  h({\bf r})   \; .
\label{hk}
\end{equation}
With $H({\bf k})$ so defined, the thermal average of MSD
in the $z$-direction is given by
\begin{equation}
  \langle h({\bf r})^2 \rangle =
	 \frac1N  \sum_{\bf k} \langle |H({\bf k})|^2 \rangle
\label{hr2}
\end{equation}
Thus, we have
\begin{equation}
  \nabla h({\bf r})  =  \frac{i}{\sqrt{N}}
     \sum_{\bf k} {\bf k} \, {\rm e}^{i {\bf k \cdot r}}  H({\bf k})
\label{nablah}
\end{equation}
and
\begin{equation}
 |\nabla h({\bf r})|^2 = \frac1N  \sum_{{\bf k}_1, {\bf k}_2} 
      {\bf k}_1 \cdot {\bf k}_2 \, 
      {\rm e}^{i ({\bf k}_1-{\bf k}_2) \cdot {\bf r}}
      H({\bf k}_1)  H({\bf k}_2)^*     \; ,
\label{nablah2}
\end{equation}
which yields
\begin{equation}
  \langle |\nabla h({\bf r})|^2 \rangle  = 
    \frac1N  \sum_{\bf k}  k^2  \langle |H({\bf k})|^2 \rangle  \; .
\label{nablah3}
\end{equation}
Then, the mean real area is given by
\begin{equation}
     A = \langle A_{\rm inst} \rangle  =  
         A_p + \frac{A_p}{2N}  \sum_{\bf k}  k^2  
        \langle |H({\bf k})|^2 \rangle  \; ,
\end{equation}
and for uncoupled vibrational modes in the out-of-plane direction
(harmonic approximation), $\langle |H({\bf k})|^2 \rangle$ can be
written as a sum of their MSDs:
\begin{equation}
  \langle |H({\bf k})|^2 \rangle = 
         \sum_j  \langle |\xi_j({\bf k})|^2 \rangle
\end{equation}
so that
\begin{equation}
 \Omega  = \frac{A - A_p}{A_p} =   \frac{1}{2N}  
     \sum_{j,{\bf k}}  k^2  \langle |\xi_j({\bf k})|^2 \rangle
\label{omega2}
\end{equation}
with
\begin{equation}
   \langle |\xi_j({\bf k})|^2 \rangle  =
       \frac{\hbar}{2 m \omega_j({\bf k})}
       \coth \left( \frac12 \beta \hbar \omega_j({\bf k}) \right)  \; .
\end{equation}
The sum in $j$ in Eq.~(\ref{omega2}) is extended to the phonon
bands with displacements in the $z$-direction
(ZA, ZO', and the two-fold degenerate ZO).

Note that in our simulations the in-plane area also fluctuates, but
its fluctuations are not considered in the harmonic calculation
presented here.

\section{Phonon contributions to the low-temperature specific heat}

Here we present a continuous model for wavenumbers and frequencies
of vibrational modes, to find an analytic dependence for the
contributions of the different phonon bands to the low-temperature
specific heat of bilayer graphene.
For a phonon branch with dispersion relation $\omega_j \propto k^n$
for small $k$, the low-temperature contribution to the specific heat
may be approximated as
\begin{equation}
        c_v^j(T) \approx   \frac{k_B}{2}   \int_{k_0}^{k_m}
    \frac { \left[ \frac12 \beta \hbar \, \omega_j(k) \right]^2 }
      { \sinh^2 \left[ \frac12 \beta \hbar \, \omega_j(k) \right] } \,
             \rho(k) \, d k  \, ,
\label{cvr}
\end{equation}
where $k_m$ is the maximum wavenumber
$k_m = (2 \pi / A_0)^{1/2}$, $A_0$ is the in-plane area for the
minimum-energy configuration, and
$\rho(k) = A_0 k / 2 \pi$ for 2D systems.
From the dispersion relation $\omega_j(k)$, we have
a vibrational density of states
\begin{equation}
        \bar{\rho}_r(\omega) =  \rho(k)  \frac{d k}{d \omega }
                 \sim \omega^{\frac2n - 1}
\label{rho_omega}
\end{equation}
so that
\begin{equation}
   c_v^j(T) \sim   k_B  \int_{\omega_0}^{\omega_m}
         \frac { ( \frac12 \beta \hbar \, \omega )^2 }
          { \sinh^2 \left( \frac12 \beta \hbar \, \omega  \right) } \,
             \omega^{\frac2n - 1}   \,  d \omega  \, .
\label{cvr2}
\end{equation}

Taking the limit $\omega_0 \to 0$ ($N \to \infty$) and
putting $x = \frac12 \beta \hbar \, \omega$, we have
\begin{equation}
   c_v^j  \sim   k_B  \frac{K}{(\beta \hbar)^{\frac2n}}
     \int_0^{x_m}  \frac { x^{\frac2n + 1} } { \sinh^2 x } \, d x   \, ,
\label{cvr3}
\end{equation}
$K$ being a constant.
At low temperature, $k_B T \ll \hbar \, \omega_m$ (large $x_m$),
we have $c_v^j \sim T^{2/n}$.
In general, for $d$-dimensional systems one has
an exponent $d/n$.\cite{po02,zi08}
Then, in graphene we expect for the ZA phonon branch ($n = 2$):
$c_v^{\rm ZA} \sim T$, and for the acoustic LA and TA branches
($n = 1$): $c_v^{\rm ac} \sim T^2$.

For the ZO' band in bilayer graphene, we have $\omega \approx \omega_0$
for small $k$. Then, at low $T$ the corresponding contribution to the
specific heat, $c_v^{ZO'}$, coincides with that of a collection of harmonic
oscillators with frequency $\omega_0$, i.e.,
$c_v^{ZO'} \sim \exp (- \hbar \omega_0 / k_B T)$.

% ----------------------------------------------------------------

%  BIBLIOGRAPHY

\end{document}